\documentclass[iop]{emulateapj}
\usepackage{verbatim}
\usepackage{mathtools}
\usepackage{natbib}
\usepackage{bm}        
\usepackage{amssymb}   
\usepackage{booktabs}

\shorttitle{On escaping a galaxy cluster in an accelerating Universe}
\shortauthors{Stark et al.}

\begin{document}
\title{On escaping a galaxy cluster \\ in an accelerating Universe}

\author{ Alejo Stark\altaffilmark{1},Christopher J. Miller\altaffilmark{1,2}  Daniel Gifford\altaffilmark{1}}
\altaffiltext{1}{Department of Astronomy, University of Michigan, Ann Arbor, MI 48109 USA}
\altaffiltext{2}{Department of Physics, University of Michigan, Ann Arbor, MI 48109, USA}
\email{alejo@umich.edu}


\begin{abstract}
We derive the escape velocity profile for an Einasto density field in an accelerating universe and demonstrate its physical viability by comparing theoretical expectations to both light-cone data generated from N-body simulations and archival data of 20 galaxy clusters. We demonstrate that the projection function  ($g(\beta)$) is deemed physically viable only for the theoretical expectation that includes a cosmology-dependent term. Using simulations, we show that the inferred velocity anisotropy is more than 6$\sigma$ away from the expected value for the theoretical profile which ignores the acceleration of the universe. In the archival data, we constrain the average galaxy cluster velocity anisotropy parameter of a sample of 20 clusters to be $ \beta = 0.248_{-0.360}^{+0.164} $ at the 68\% confidence level. Lastly, we briefly discuss how our analytic model may be used as a novel galaxy cluster-based cosmological probe. 

\end{abstract}

\keywords{ cosmology: theory --  cosmology: observations -- galaxies: clusters: general }


\section{Introduction}
\label{sec:intro}
The discovery of the late time acceleration of the universe is one of the most profound mysteries of physical cosmology. What is at stake with this discovery is the following: either our universe is composed of some exotic "dark energy"  whose physics drives the dynamics of acceleration and/or our general relativistic theory of gravity must be extended or modified \citep{joyce,koyama2016, joyce2016}.

Given its profound importance, cosmic acceleration is currently being investigated through a broad constellation of probes that mobilize a range of astrophysical objects and phenomena such as: Type Ia supernovae,  Baryon Acoustic Oscillations (BAO), weak gravitational lensing and galaxy clusters (see \cite{weinberg2013} for an excellent review of these and other approaches).

Galaxy clusters in particular are vital laboratories that allow us to sensitively probe the physics of large-scale structure formation and thereby constrain cosmological models of our universe \citep{kravtsovborgani2012}. The method most commonly applied is based on the cluster abundance function, which evolves in shape and amplitude as a function of the cosmological parameters \citep{vikhlinin,rozo2010}. The  abundance function as a cosmological probe depends not only on robust analytic theory which is calibrated through simulations, but also on accurate masses as inferred from dynamical, weak lensing and X-ray methods.

An alternative way of constraining cosmology that does not depend directly on the galaxy cluster abundance function was developed in \cite{regos} and extended by \cite{regoes1996}. Both of these papers constrain the cosmological matter density parameter through an analysis of the phase (velocity vs. distance) space of galaxy clusters. More specifically, as demonstrated by \cite{kaiser1986} the infall pattern around galaxy clusters forms a trumpet-shaped profile in their phase spaces. This trumpet-shaped profile, also known as a phase space ``caustic'', can be inferred from the line-of-sight velocity information. When compared to what is predicted by spherical infall models, an estimate on the matter density parameter of the universe may be inferred. However, the capacity for the caustic amplitude and shape to precisely constrain the matter density parameter, at least when considering both linear and non-linear theory from the spherical infall model, have since been called into question \citep{diaferiogeller1997}.

In particular, \cite{diaferiogeller1997} demonstrated that the caustic profiles were related to the escape velocity profile of the cluster as mediated by a projection function. This projection function stems from the fact that we observe only the line-of-sight component of a galaxy's velocity and because the true velocity vectors can be non-isotropic. Only after projection can the caustics be utilized to infer the mass profiles of galaxy clusters. While this method opened the path for a novel way of estimating the mass profiles of galaxy clusters, the capacity for the caustic to constrain cosmology directly has not been pursued. 

In what follows, we argue that in order to properly model the escape velocity profile of galaxy clusters as inferred from their phase spaces a cosmological acceleration term must be included. The reason for this is that the escape velocity profile is often defined by integrating the density profile out to infinity via the integral form of the Poisson equation. However, current analytical expressions of the gravitational potential of galaxy clusters at "infinity" are not well defined. As such, if the potential is not properly normalized it will yield the wrong escape velocity profile.

A recent analysis by \cite{miller2016} utilized 3-dimensional phases spaces from the Millennium simulation \citep{springel2005millenium} to demonstrate that the escape velocity edge (e.g., the 3-dimensional caustic) can recover the true underlying escape velocity profile of galaxy clusters to high accuracy and precision. The true escape velocity profile is determined from the application of the Poisson equation to the measured cluster density profile. As we discuss below, this technique requires cosmological information pertaining to the  acceleration of space.

In their analysis, \cite{miller2016}  also report that the NFW model overestimates the escape velocity profiles of galaxy clusters by $\sim 10\%$ \citep{navarro96, navarro97}. This is because the NFW density profile over-estimates the mass beyond the virial radius. Other analytic models of the density profile of dark matter halos, such as the Einasto \citep{einasto65} and the Gamma model \citep{gamma}, fare much better. We note that each of these analytical representations of the density profile can be constrained to be identical within $\sim r_{200}$ (the radius at which the average density drops to $200\times$ the critical density of the universe). Where these density profiles differ is in the outskirts, where both the Einasto and the Gamma profiles are steeper than the NFW profile.

In this effort, we extend the work of \cite{miller2016} to include projection effects and also to test the theory and the algorithm on real data. We use projected synthetic data from the \cite{henriques} light-cone as well as archival data of 20 galaxy clusters with extensive spectroscopic data and weak lensing mass profiles. We analyze the cosmology-dependent escape velocity profiles as inferred from their projected phase spaces and assess the viability of our analytic expectations. We note that only when including cosmological effects do we recover values for the velocity anisotropy parameter that are in agreement with $\Lambda$CDM simulations and with various other observational studies: \cite{lokasabell},  \cite{benatov}, \cite{lemze2009}, \cite{wojtak}, and \cite{macs1206}. 

The outline of our paper is as follows: in Section 2 we derive and review the theoretical expectations related to our observable: the escape velocity profile of galaxy clusters. In Section 3 we outline how this observable is inferred from the phase space of galaxy clusters. From thereon, in Section 4, we describe both the synthetic and non-synthetic projected data we utilized to probe our theoretical expectations. In Section 5 we describe projection effects and estimate the value of the velocity anisotropy parameter we infer assuming two different theoretical expectations of the escape velocity profile: one that includes the cosmological term and one that does not. In the following section, Section 6, we thoroughly assess likely sources of systematics affecting our analysis. We follow this analysis with a discussion and conclusion in sections 7 and 8 respectively.

Except for the case of synthetic data in which the cosmological parameters are already defined \citep{springel2005millenium}, in what follows we assume a flat $\Lambda$CDM cosmology with $\Omega_{M}= 0.3$, $\Omega_{\Lambda}=1-\Omega_{M}$, and $H_0 = 100 h \text{ km}\text{ s}^{-1}\text{ Mpc}^{-1} $ with $h = 0.7.$

\section{Theoretical expectations}
The theory of general relativity and its derivative cosmological models demonstrate that the dynamics of the matter-energy and the universe's expansion dynamics are dialectically entwined. As it is often said: matter-energy tells space-time how to curve and space-time tells matter-energy how to move. For instance, in the case of galaxy cluster-sized halos, large scale cosmological dynamics are expressed in the amplitude and shape of the halo mass function \citep{tinker}. Qualitatively, what this tells us is that the dynamics of the galaxies in a gravitationally bound structure such as a galaxy cluster must also necessarily be affected by cosmology. 

In this paper we focus on a particular observable, the escape velocity profile of clusters as inferred from their phase space, and test the ways in which analytical formulations of this observable must necessarily introduce a cosmological term in order to accurately describe the escape velocity edges of galaxy clusters. In particular, the velocity profile ($v_{esc}$) can be inferred analytically by characterizing the potential ($\phi$) a given test particle must escape from,

\begin{equation} 
v_{esc}^2 = -2\phi.
\label{eq:vesc_phi}
\end{equation} 

However, as mentioned in section \ref{sec:intro}, the cosmological effect on the escape velocity profile has to be included in this potential. We now derive the escape velocity profile that includes cosmology and which has been utilized and tested against both  $\Lambda$CDM universe simulations \citep{behroozi, miller2016} and extensions to general relativity such as Chameleon $f(R)$ gravity \citep{starkmiller}. We then extend those derivations to include projection effects.

\subsection{Escape velocity profile of a galaxy cluster \\ in an accelerating universe}
\cite{nandra} demonstrated that in the weak field approximation of general relativity and at sub-horizon scales, a massive particle will still feel a force from the accelerated expansion of space. Following \cite{nandra}, then, the effective acceleration experienced by a massive particle with zero angular momentum in the vicinity of a galaxy cluster with gravitational potential ($\Psi$) is given by,

\begin{equation} 
\vec\nabla \Phi =  \vec\nabla \Psi + qH^2 r \hat{r}.
   \label{eq:acceleration_eq}\end{equation} 
The effective potential ($\Phi$) therefore takes into account both the curvature produced by a density field with potential $\Psi$ and the curvature produced by the acceleration term $qH^2r$. From a Newtonian perspective, then, this last term can be thought of as a repulsive force that opposes the inward pull of the cluster's mass distribution.

In the second term of equation \ref{eq:acceleration_eq}, $q$ is the usual deceleration parameter given by: $q\equiv  -\frac{\ddot{a}a}{\dot{a}^2}$.We assume a flat universe ($\Omega_k=0$), and a dark energy equation of state parameter $w=-1$. Also, given that we  work in low-redshift regime (such that $\Omega_{\gamma}(z) \approx 0 $) the deceleration parameter can be expressed in terms of the redshift evolving matter density parameter ($\Omega_M(z)$) and the redshift evolving $\Lambda$ density parameter ($\Omega_{\Lambda}(z)$), $ q(z) = \frac{1}{2} \Omega_{M}(z)  - \Omega_{\Lambda}(z).$ For our chosen cosmology at the present epoch, we attain: $q(z=0) = -0.55.$ Lastly, the Hubble parameter ($H$) for this same cosmology is as usual, $H(z) = H_0 E(z) = H_0 \sqrt{ (1- \Omega_{M}) + \Omega_{M} (1+z)^3 }$.

Having defined the cosmological quantities that compose equation \ref{eq:acceleration_eq}, we can integrate over the physical radius ($r$) to find the effective potential, and subsequently the escape velocity profile via equation \ref{eq:vesc_phi} with effective potential $\Phi$. Integrating equation \ref{eq:acceleration_eq},

\begin{equation} 
\int_r^{r_{eq}} d\Phi =  \int_r^{r_{eq}}  d\Psi + qH^2 \int_r^{r_{eq}}  r' dr'.
   \label{eq:acceleration_eq_int}\end{equation} 
Note that we are integrating out not to infinity but to a finite radius, $r_{eq}$, which is termed the ``equivalence radius'' in \cite{behroozi}. The reason for this is that the escape velocity at infinity is poorly defined. This ambiguity introduces a problem with the normalization of the potential that is used to calculate the escape velocity profile. In particular, as demonstrated in \cite{miller2016} this offset ends up overestimating the potential by $\sim$20\%. 
Following \cite{behroozi} we define the equivalence radius to be the point at which the acceleration due to the gravitational potential of the cluster  and the acceleration of the expanding universe are equivalent ($\vec\nabla \Phi= 0$) which yields, $r_{eq} = ( \frac{GM}{-qH^2} )^{1/3},$ where $G$ is the gravitational coupling constant and $M$ is the mass of the cluster as inferred from a choice of $\Psi$ via the Poisson equation, as detailed in subsection 2.2 below. Now, integrating  equation \ref{eq:acceleration_eq_int} out to $r_{eq}$, we have,
 \begin{equation} 
\Phi(r) =  \Psi(r) - \Psi(r_{eq})  + \frac{1}{2} qH^2 \big( r^2- r_{eq}^2 \big) + \Phi(r_{eq}).
\end{equation} 
Setting the boundary condition such that the escape velocity must necessarily be 0 at the equivalence radius, $-2\Phi(r_{eq}) = v_{esc}^2(r_{eq}) = 0$, and using equation \ref{eq:vesc_phi} we find,

\begin{equation} 
v_{esc}(r) = \sqrt{-2 \big(\Psi(r) - \Psi(r_{eq}) \big) - qH^2 \big(r^2 - r_{eq}^2 \big)}.
 \label{eq:vesc}
\end{equation}
This reproduces the result shown in \cite{starkmiller} and \cite{miller2016}. From now on we refer to this escape velocity profile as Einasto $qH^2$.

Note that equation \ref{eq:vesc} is  therefore normalized to yield an escape velocity of zero at the equivalence radius once a $\Psi$ has been chosen. Note also that equation \ref{eq:vesc} yields the escape velocity profile in an accelerating universe for any choice of gravitational potential $\Psi$.

\subsection{Gravitational potential}
\label{sec:theory_pot}
While it is common to describe the density profile of galaxy clusters with the NFW model of dark matter halos \citep{navarro96, navarro97}, recent investigations have shown that the NFW potential-density pair over estimates the escape velocity profile by $\sim$10\% within galaxy clusters \citep{miller2016}. This is because, on average, the NFW profile tends to overestimate the mass in the outskirts of galaxy clusters (see also \cite{Diemer2015}).

Moreover, as is also demonstrated by \cite{miller2016}, in contrast to the NFW model, once the cosmological term (Eq. \ref{eq:vesc}) has been included, both the Gamma \citep{gamma} and Einasto \citep{einasto65} gravitational potential profiles can model the radial escape velocity profiles to better than $3\%$ precision. In what follows, we utilize the Einasto profile. However, we emphasize that our analysis is not dependent on the $\Psi$ used. That is, in so far as the gravitational potential profiles are derived from a true density-potential pair, all of our analysis will yield the same results.

Because nearly all published weak-lensing mass profiles utilize the NFW model, we need to map the NFW parameters to the equivalent parameters in the Einasto profile. The Einasto representation of dark matter halos \citep{einasto65} is a three parameter model ($n, \rho_{0}, r_0$) described by the following fitting formula for the density profile,
\begin{equation}
\rho(r) =  \rho_0 \rm{exp} \Bigg[ -\bigg(\frac{r}{r_0}\bigg)^{1/n} \Bigg].
\label{eq:einasto_den}
\end{equation}

From the density field described by equation \ref{eq:einasto_den}  we can derive the gravitational potential using the integral form of the Poisson equation. As demonstrated by \cite{retana} this yields,

\begin{equation}        
\Psi(r) = -\frac{{\rm GM}}{r} \Bigg{[} 1 - \frac{\Gamma\big{(}3n,\big(\frac{r}{r_0}\big)^{1/n}\big{)}}{\Gamma(3n)} + \frac{r}{r_0}\frac{\Gamma\big{(}2n,\big(\frac{r}{r_0}\big)^{1/n}\big{)}}{\Gamma(3n)}\Bigg{].} \label{eq:einasto_pot}
\end{equation}

As in \cite{miller2016}, $\rho_0$ (or the mass term $M$) can be thought of as the normalization, $r_0$ is the scale radius and $n$ is the index. Following \cite{retana}, we use $\Gamma(3n) = 2\Gamma(3n, d_{n})$ where the $d_n$ term is solved for via asymptotic expansion. In particular, we use the $d_{n}$ term cited therein expanded up to the fifth order. Lastly, $\Gamma(a,x)$ denotes the upper incomplete gamma function, given as usual by: $\Gamma(a,x) = \int_x^{\infty} t^{a-1} e^{-t} dt.$

As shown in \cite{sereno2016} and references therein, the mapping between NFW and Einasto profiles is straightforward. In our case, we insist that the two profiles be nearly identical within $r_{200}$. We do this by solving for the Einasto parameters which match analytical NFW density profiles on a cluster-by-cluster basis. As noted by \cite{sereno2016}, recent weak lensing analyses of stacked clusters cannot distinguish between the Einasto and NFW halo representations of the density profile within this range. 

\subsection{Comparing theoretical escape velocity profiles \\ with and without the cosmological terms}

\begin{figure}
\epsscale{1.2}
\plotone{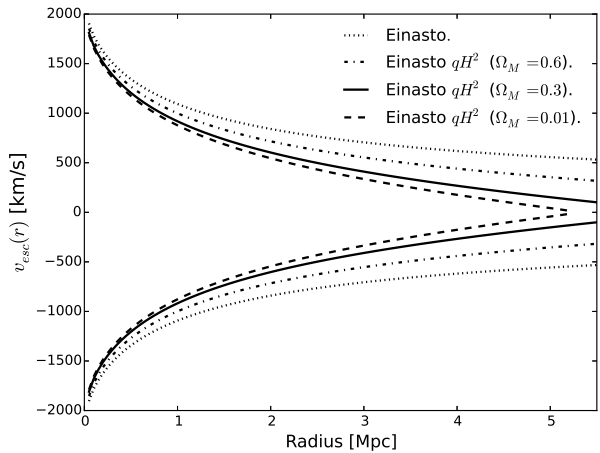}
\caption{Escape velocity profiles for a single cluster of fixed mass 
 using equations \ref{eq:vesc} and \ref{eq:einasto_pot} for four different escape velocity profile models ($v_{esc}(r)$). Note that while we change $\Omega_M$ we keep $h = 0.7$ for a fiducial flat $\Lambda$CDM cosmology. We note the significant difference between the Einasto $qH^2$ theoretical expectations for three different flat $\Lambda$CDM cosmologies with differing matter density parameters ($\Omega_M = 0.01, 0.3, 0.6$, denoted by dashed, solid, and dashed dotted lines respectively) and the "Einasto" theoretical expectation which does not include a cosmological term (denoted by dotted line) given by $v_{esc}(r) = \sqrt{-2\Psi(r)}$ (and using equation \ref{eq:einasto_pot}). Note that increasing $\Omega_M$ raises the escape velocity at all radii. Going in the opposite direction, we notice that as our universe becomes more and more dominated by $\Lambda$ it will in principle be easier for galaxies to escape clusters. Secondly, we note that as $\Omega_M$ increases we recover the non-cosmological escape velocity (dotted line). Note also, as implied by equation \ref{eq:vesc}, that the Einasto $qH^2$ profiles all eventually reach some radius (the equivalence radius "$r_{eq}$") at which the escape velocity is zero.}
 \label{fig1}\end{figure}

In Fig. \ref{fig1} we show the resulting profiles for a single cluster-sized halo at $z=0$. In particular we plot both the escape velocity profiles for the Einasto $qH^2$ profile (Eq. \ref{eq:vesc}) with three different cosmologies (dashed, solid and dash-dotted lines) and for the Einasto profile without the cosmological term  (dotted line): $v_{esc}(r) = \sqrt{-2\Psi(r)}.$

As set by our aforementioned boundary condition, the Einasto $qH^2$ escape velocity profiles all reach an equivalent radius where $v_{esc}(r) \rightarrow 0$. We also highlight the significant difference between the Einasto $qH^2$ model with $\Omega_M = 0.3$ and the Einasto model without the cosmological term. Lastly, note that as we increase $\Omega_M$ our Einasto $qH^2$ profiles converge to the Einasto profile without the cosmological term.

Given that all other data indicates we do not live in an Einstein-de Sitter universe, we should be able to detect the $\gtrsim 10\%$ cosmology-dependent effects on the escape velocity profiles of galaxy clusters. We first test our theoretical expectations in N-body simulations in order to thoroughly assess systematic effects (including projection effects, mass scatter, and more, as explained in the subsequent sections) and then utilize archival redshift data and weak lensing mass estimates of 20 galaxy clusters to test our expectations.


\section{Observables}
As is clear from our theoretical expectations, our observables are the projected escape velocity profile of galaxy clusters ($v_{esc}^{edge}(r)$) and the observed weak-lensing mass profile. We use the latter in our analytic model of the 3-dimensional escape edge and we require the cosmology and the projection function as described in Section \ref{sec:infer_beta}.

We utilize the redshift information of galaxy clusters to generate phase spaces ($v_{los}$ vs. $r$ space) from which we infer the escape velocity profile. More specifically, we do this by transforming the galaxy redshifts at a given angular separation from the cluster center ($\theta$) to line-of-sight ($v_{los}$) velocities at the cluster's redshift ($z_c$)  via,
\begin{equation} 
v_{los} =  c\frac{(z - z_c)}{{(1+z_c)} }.
\end{equation} 
Where $c$ is the speed of light, and  $z$ denominates the redshift of individual galaxies. To calculate the physical distance from the cluster's center ($r$) we calculate the angular diameter distance ($d_A$) and use the angular separation ($\theta$), 
\begin{equation} 
r = d_{A}(z) \theta = \bigg[\frac{1}{1+z} \frac{c}{H_0} \int_0^z \frac{dz'}{E(z')}\bigg] \theta.
\end{equation} 
For each cluster phase space we then identify the edge in radial bins of 0.1 Mpc by finding galaxies with the top 10\% velocities. We follow the interloper removal prescriptions of \cite{gifford2013} and Gifford et al. (2016). The latter tested this edge detection technique on projected data in two different simulations with widely varying values for $\sigma_8 = 0.8$ vs $\sigma_8 = 0.9$, where $\sigma_8$ is the normalization of the matter power spectrum on 8 Mpc scales. They conclude that the edge detection is independent of large variations in the line-of-sight interloper fraction.  We also test the robustness of this edge-detection technique as discussed in the Systematics section below.
There are various techniques used to define phase space edges in the literature \citep{diaferiogeller1997,gifford2013, serra, hecs,geller2013,lemze2009,miller2016}. With N-body simulations \cite{miller2016} have shown that the escape velocity edge ($v_{esc}^{edge}$) can be inferred with $\sim5\%$ accuracy. We discuss all of this more thoroughly in the Systematics section.

The weak lensing mass profiles are taken from the literature \citep{sereno}. As noted in Section \ref{sec:theory_pot}, these are provided as NFW profiles and we convert them to Einasto density profiles on an individual basis. When the NFW concentration parameter is not provided, we use the mass-concentration from \cite{duffy}.  We include the weak lensing mass errors as provided in the literature.



\section{data}
We first test our theoretical expectations with a sample of synthetic clusters from the\cite{henriques} light-cone data. After having assessed relevant systematics (thoroughly described in Section 6 below) with the synthetic data we conduct our analysis on 20 clusters with weak lensing and redshift data. We briefly discuss these data sets below.

\subsection{Synthetic data}
We utilize the Millenium simulation  \citep{springel2005millenium} which employs a flat cosmology with $\Omega_M = 0.25$ and $h= 0.73$. In particular we pick all the clusters above $ M_{200} > 4 \times 10^{14} h^{-1} M_{\odot}$ in the \cite{henriques} light-cone and from there we ensure that each cluster has $N \gtrsim 110$ galaxies within $2 r_{200}$ and between $-2000 \leq v_{los} \leq 2000 $ km/s. We do this to ensure that the phase spaces we are working with are well sampled so that we may accurately infer escape velocity edges from them. From these cuts, we end up working with a sample of 200 halos. We then cross-correlate this light-cone data sample with the \cite{guo} catalog in order to obtain 3-dimensional velocity information for each of our projected clusters in the \cite{henriques} light cone. This 3-dimensional information is needed in order to compute the velocity anisotropy parameter of each cluster. We expand on this in Section 5 below. Out of the sample of 200 clusters we separate them into ten sets of 20 with similar mass distributions in order to create a sample with comparable statistics as our archival non-synthetic galaxy cluster sample.

\subsection{Archival data: \\ weak lensing masses and galaxy redshifts}
We used the VizieR catalog \citep{vizier} to search for redshift data of galaxy clusters with weak lensing mass estimates. The galaxy redshift information is taken from a variety of sources  \citep{hecs,millerabell,ellingson1997,sanchez,geller2014,CLASHvlt,owers2011,CAIRNS,lemze2013}
and so are the weak lensing mass estimates \citep{okabe2008,okabe2015,hoekstra2015,umetsu2012,gavazziComa}. The references for both weak lensing mass profiles and galaxy redshifts for each of the 20 clusters in our sample are summarized in Table 1. Note that while we cite the original source of the weak lensing papers we ultimately use the $M_{200}$ masses and errors in our analysis  compiled in the \cite{sereno} meta catalog. We particularly chose to utilize this meta catalog because it includes standardized weak lensing mass estimates across cosmologies.\footnote{The latest and most updated weak lensing parameter estimates from the meta catalog can be accessed through Mauro Sereno's website: \url{pico.bo.astro.it/~sereno/CoMaLit/LC2/2.0/}.} Note that the only case in which we use the exact mass as made explicit in the weak lensing papers listed in Table 1 is for A1656 given that this cluster is not included in the \cite{sereno} meta catalog.

\begin{table}[]
\centering
\caption{List of Galaxy Clusters and References}
\label{table1}
\begin{tabular}{@{}llll@{}}
\toprule
Cluster name\footnote{While we cite the original papers above, the weak lensing masses (and their respective errors) we use in our analysis were taken from the \cite{sereno} meta catalog. More specifically, \cite{sereno} standardizes the $M_{200}$  masses for the clusters shown above (as inferred from each reference listed in the "weak lensing" column) for the fiducial cosmology mentioned in our introduction.} & Redshift & Weak lensing\footnote{The abbreviations in this column refer to the following papers: H15= \cite{hoekstra2015}, OU08 = \cite{okabe2008}, OS15= \cite{okabe2015}, U12= \cite{umetsu2012}, G09 = \cite{gavazziComa}.} & Galaxy redshifts \\ \midrule
A2111 & 0.229 &  H15 & Miller et al. '06 \\
A611 & 0.288 & H15 & Lemze et al. '13 \\
MS1621 & 0.428 & H15 & Ellingson et al. '97 \\
Cl0024 & 0.3941 & H15 & Sanchez et al. '15 \\
A2259 & 0.164 & H15 & Rines et al. '13 \\
A1246 & 0.1921 & H15 & Rines et al. '13 \\
A697 & 0.2812 & H15 & Rines et al. '13 \\
A1689 & 0.1842 & H15 & Rines et al. '13 \\
A1914 & 0.166 & H15 & Rines et al. '13 \\
A2261 & 0.2242 & H15 & Rines et al. '13 \\
A1835 & 0.2506 & H15 & Rines et al. '13 \\
A267 & 0.2291 & H15 & Rines et al. '13 \\
A1763 & 0.2312 & H15 & Rines et al. '13 \\
A963 & 0.206 & H15 & Rines et al. '13 \\
A383 & 0.187 & H15 & Geller et al. '14 \\
A2142 & 0.0909 & OU08 & Owers et al. '11 \\
RXJ2129 & 0.2339 & OS15 & Rines et al. '13 \\
A2631 & 0.2765 & OS15 & Rines et al. '13 \\
MACS1206 & 0.439 & U12 & Zitrin et al. '12 \\
A1656 & 0.0237 & G09 & Rines et al. '03 \\ \bottomrule
\end{tabular}
\end{table}

As with the synthetic galaxy cluster sample, all 20 of our clusters have  $N \gtrsim 110$  galaxies within $2 r_{200}$ and between $-2000 \leq v_{los} \leq 2000 $ km/s. The only exception is A2111 which has $N = 87$ galaxies within that range. The mass range of the data lies between $5\times10^{14}M_{\odot}$ and $2.6\times10^{15}M_{\odot}$.

The meta catalog only lists masses inferred from NFW fits \citep{navarro96, navarro97} to weak lensing shear measurements. As discussed in Section \ref{sec:theory_pot}, we convert the NFW profiles to Einasto density profiles. 


\section{testing our theoretical expectations}
To summarize, with galaxy redshift information for each cluster we  create a phase space with its line-of-sight velocities ($v_{los}$). From this phase space we infer the escape velocity edge ($v_{esc}^{edge}$), as detailed in Section 3. On the other hand, with mass profile measurements we generate an analytic escape velocity profile through equation \ref{eq:vesc}  after assuming a gravitational potential with the form of equation \ref{eq:einasto_pot} (i.e. the Einasto model of dark matter halos). We expect that if our theoretical expectations can reproduce the edge profile to high precision then the average ratio $v_{esc}^{edge}/ v_{esc}$ should yield unity. What \cite{miller2016} has shown is that the Einasto model with the additional cosmological term can  analytically reproduce velocity escape edges inferred from 3-dimensional phase spaces to high precision. Therefore, when comparing our analytic formulation with edges inferred from projected phase spaces, any difference between this ratio and unity should arise from projection effects.

\subsection{Projection effects}
As demonstrated by \cite{diaferiogeller1997} and \cite{diaferio1999} the 3-dimensional escape velocity profile ($v_{esc}(r)$) can be projected by a function of the velocity anisotropy parameter ($g(\beta)$), 

\begin{equation} 
v_{los}(r) = v_{esc}(r) \times \big({\sqrt{g(\beta(r))}}\big)^{-1}.
\end{equation} 

Where $g(\beta(r))$ is given by,
\begin{equation} 
g(r) = \frac{3 - 2\beta(r)}{1-\beta(r)}.
\label{eq:gbeta}\end{equation}

The anisotropy parameter ($\beta$) is given by the ratio of the velocity dispersion in the tangential direction ($\sigma_t$) to the velocity dispersion in the radial direction ($\sigma_r$),

\begin{equation}  
\beta(r) = 1 - \frac{\sigma_{t}^2}{\sigma_{r}^2}.
\label{eq:beta}\end{equation} 

Where $\sigma_{t}^2 = \frac{1}{2} \big(\sigma_{\theta}^2 +\sigma_{\phi}^2 \big) $ includes both azimuthal and latitudinal velocity dispersions. The limiting cases are:  radial infall ($\beta=1$),   circular motion ($\beta = -\infty$) and isotropy ($\beta = 0$). 

Therefore, if we had the 3-dimensional velocity information for each of our clusters we can calculate $\beta$ and directly project our profiles. However, in practice this parameter is difficult to determine and as such we are left to infer what $\beta$ would be given an expected theoretical profile for a given cluster and compare our result with simulations. 

\subsection{Inferring the anisotropy parameter}
\label{sec:infer_beta}
Given that our theoretical expectation matches the edges in 3-dimensions to high precision, by taking the ratio between the escape velocity edge and our theoretical profile we should be able to infer the anisotropy parameter via,
\begin{equation} 
\bigg \langle \frac{v_{esc}^{edge}(r)}{v_{esc}(r)}  \bigg  \rangle \big\langle {\sqrt{g(r)}} \big\rangle =1.
\label{eq:ratio}
\end{equation}
The brackets in equation \ref{eq:ratio} signify that the average is calculated on $N=20$ clusters. That is, we calculate the ratio for each cluster and then take the average at each radial bin, with a separation of $\Delta(r/r_{200}) = 0.1.$  Moreover, the averaged ratio is weighted by the error on the ratio of each individual cluster at a given radial bin. The details of our error budget are thoroughly discussed in the Systematics section below.

Using equation \ref{eq:ratio}, for a given average ratio, we can find what average $g(\beta(r))$ is needed to make that ratio unity. Then by inverting equation \ref{eq:gbeta} we can find the anisotropy parameter $\beta$ via,
\begin{equation} 
\beta(r) = \frac{3-g(r)}{2-g(r)}.
\label{eq:betafromg}
\end{equation} 

Lastly, the theoretical expectation ($v_{esc}$) can be either the escape velocity profile with the cosmological term (i.e. Einasto $qH^2$, Eq. \ref{eq:vesc}) or without it (i.e. just the Einasto potential, Eq. \ref{eq:einasto_pot}: $v_{esc}(r) = \sqrt{-2\Psi(r)}$). The differences between the inferred $\beta$ parameters for these two analytic profiles are detailed in section 5.2.1 and 5.2.2 for the synthetic sample, respectively. For the archival data we only compare our inferred $\beta$ assuming the Einasto $qH^2$ theory (see section 5.2.3). 

\subsubsection{Synthetic data and theory with cosmological term}
\begin{figure}
\epsscale{1.2}
\plotone{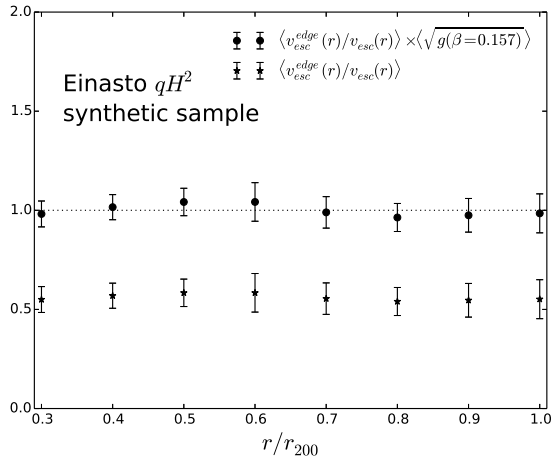}
\caption{The unprojected (stars) and projected (circles) weighted average ratio between the escape velocity edge and the theoretically expected escape velocity profiles with the Einasto $qH^2$ model for a single set of 20 synthetic clusters from the \cite{henriques} light cone. The error is the 1$\sigma$ error on the averaged ratio assuming a uniform 25\% mass scatter on the virial mass of each cluster and a 5\% scatter on the escape velocity edge. The unprojected ratios are projected after calculating the most likely average $\beta$ by comparing to the unity line (see Equation \ref{eq:ratio}). We find this value to be well in agreement with simulation results (see figure \ref{fig3}).}
\label{fig2}
\end{figure}

Assuming $v_{esc}(r)$ in equation \ref{eq:ratio} is the Einasto $qH^2$ model (Eq. \ref{eq:vesc}), for a single set of 20 clusters in our synthetic sample, the weighted average ratio is shown in Fig. \ref{fig2} (black stars). 

With equations \ref{eq:ratio} and \ref{eq:betafromg} we then calculate the $\chi^2$ between $0.3 < r/r_{200} < 1$ in order to infer the most likely average $\beta$ for this set of 20 clusters. Note that we focus on the region between $0.3 < r/r_{200} < 1$ because simulation results have shown that the anisotropy parameter is on average constant across different redshifts within this radial range \citep{serra,lemze,munari}. 

After inferring $\beta$ from this method we project the average profile. The results for this sample are also shown in Fig. \ref{fig2} (black dots).

\begin{figure}
\epsscale{1.2}
\plotone{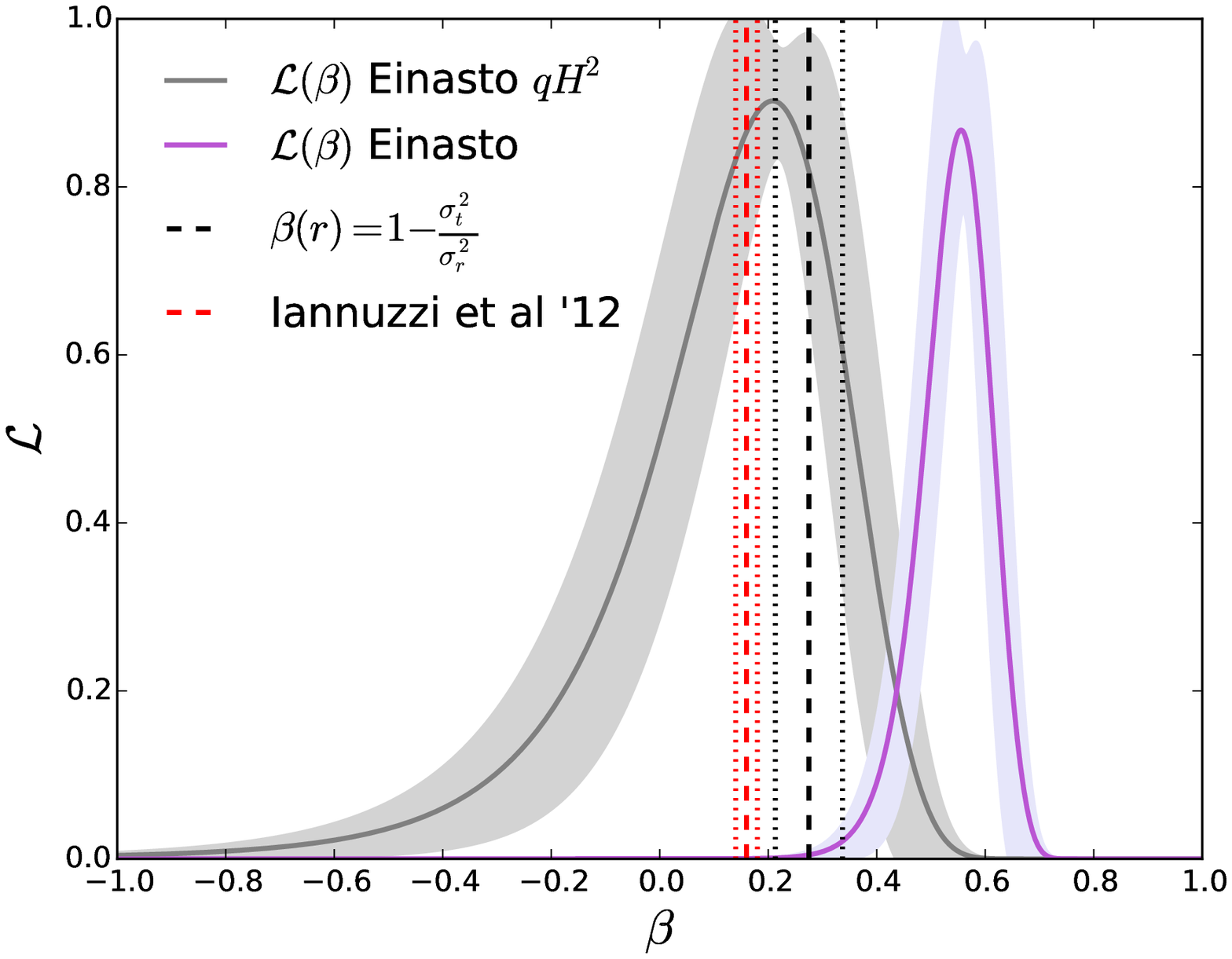}
\caption{The likelihood of the average anisotropy parameter ($\beta$) as inferred from ten sets of 20 clusters in the synthetic \cite{henriques} light-cone data. The gray band represents the $1\sigma$ variation on the likelihood of the ten sets.  We cross-correlated the \cite{henriques} light-cone data with the 3-dimensional velocity data from \cite{guo} to calculate, analytically, the average beta for the sample via Eq. \ref{eq:beta} to attain: $\beta = 0.275 \pm 0.062 $ at the 95\% confidence level (black line dashed and dotted). The red lines represent the $\beta$ profile with 2$\sigma$ error (dash and dotted) for 1000 clusters in the Millenium sample calculated also with Eq. \ref{eq:beta} by \cite{iannuzzi}. Only by including the cosmological-dependent term on our theoretical formulation can we recover the true $\beta$  with accuracy and to high precision. Compare the red lines to the purple band which utilizes the Einasto potential for the analytic profile of the escape velocity profile without the cosmological term. We conclude, then, that we can rule out the analytic profile without the cosmological term at the $6.3\sigma$ level. Note that the \cite{iannuzzi} band is the error on the mean as calculated from a bootstrap algorithm also between $0.3 < r/r_{200} < 1$. Lastly, as mentioned in the text, we assume a uniform 25\% mass scatter on $M_{200}$ for all clusters and a 5\% error on the escape velocity edge. The cosmology utilized for our analytic profiles is the same as what was utilized to make the simulations (see section 4).}
\label{fig3}
\end{figure}

Moreover, with the $\chi^2$ we calculated, and assuming a Gaussian likelihood $\mathcal{L} \propto \exp[ -\chi^2/2 ]$, we can generate a likelihood plot of our inferred average $\beta$ for each of our ten sets of 20 synthetic clusters. The result is shown by the gray band in Fig. \ref{fig3}. The band represents the 1$\sigma$ error on the distribution of likelihoods for all ten sets of averaged cluster ratios and their inferred $\beta$'s. In that same figure we compare our inferred $\beta$ with two other results from synthetic data. 

The red vertical dashed line is the average anisotropy parameter (also between $0.3 < r_{200} < 1$) as directly measured in 3-dimensions using Eq. \ref{eq:beta} calculated by \cite{iannuzzi}. The $2\sigma$ bootstrap error on the mean is shown in red dotted lines. The sample used by \cite{iannuzzi} is composed of the 1000 clusters in the Millennium simulation at $z=0$ with virial masses greater than $ 2 \times 10^{14} M_{\odot}$. This is a much larger sample size than we use and as such we consider this $\beta$ to be a robust estimate of the true anisotropy parameter. 

The black vertical dashed lines in Fig. \ref{fig3} come from a direct calculation of the anisotropy parameter (using Eq. \ref{eq:beta}) for the superset of 200 clusters with masses $M_{200} > 4\times10^{14}M_{\odot}$ that we use in this work. 
Our superset has a slightly larger average $\beta$, but both our measurement and \cite{iannuzzi}'s are within $\sim 2\sigma$ of each other. We hypothesize that this small difference could be attributed to the fact that our simulated data include the orphan galaxies in the \cite{guo} catalog, whereas \cite{iannuzzi} exclude these. Note, for instance, that the $\beta$ calculated in \cite{lemze} utilized particles and is also larger than \cite{iannuzzi}'s.

The gray line (and band) is the likelihood which represents $\beta$ using only the projected phase-space profiles of 20 clusters (i.e., no 3-dimensional information). As expected, this likelihood is much larger than the error bounds on $\beta$ from the 3-dimensional information for larger samples. The reason it is larger is because the $\chi^2$ analysis includes representative errors on both the weak lensing masses (25\%) and the escape edges (5\%). However, this likelihood fully captures the true underlying 3-dimensional radially averaged velocity anisotropy.   

\subsubsection{Synthetic data and theory without cosmological term}

In both of our synthetic data samples shown in Fig. \ref{fig2} and Fig. \ref{fig3} we took the ratio between escape velocity edge and the Einasto $qH^2$ analytic profile. Now, assuming we do not actually need the cosmological-dependent term to accurately reproduce the escape edge, we test whether or not we can recover the true anisotropy parameter. More specifically, the $v_{esc}(r)$ in the ratio now utilizes Eq. 1 with the Einasto potential of \ref{eq:einasto_pot}: $v_{esc}(r)=\sqrt{-2\Psi(r)}$ (rather than Eq. \ref{eq:vesc}). The result is shown in Fig. \ref{fig3} (purple band). 

As with the gray band described above, the purple band represents the likelihood on the inferred $\beta$ with the $1\sigma$ error representing the standard deviation in the ten sets of 20 synthetic clusters. We see that if we remove the cosmological term, the average anisotropy parameter is much larger and therefore we cannot  recover the simulation results. More specifically, assuming the red bands are correct in Fig. \ref{fig3}, that is, if it truly describes the anisotropy parameter, then we can rule out the non-cosmological dependent theory (purple band) at the $6.3\sigma$ level.

In summary, these tests on the simulations provide two important results: first that our algorithm can recover the average $\beta$ given only projected data and second that the cosmological model plays a significant role.

\subsubsection{Archival data and theory with cosmological term}

After getting a sense of what we should expect for a given sample size of 20 clusters, and after making concrete the abstract necessity of including a cosmological dependent term in our escape velocity profile by studying the underlying velocity anisotropy distributions with 200 synthetic clusters, we perform the same analysis on an archival data set of 20 galaxy clusters. 

\begin{figure}
\epsscale{1.2}
\plotone{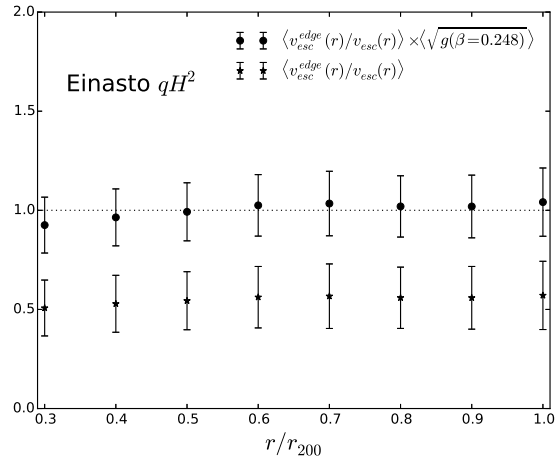}
\caption{The unprojected and projected weighted average ratios between the escape velocity edge and the theoretically expected escape velocity profiles for the set of archival data set of 20 clusters. Note the similarity between this sample and the synthetic sample Fig. \ref{fig2}. Given that we are using the particular weak lensing mass errors for each cluster (rather than a uniform mass scatter of 25\% as in  Fig. \ref{fig2}) the overall error on the average ratio is larger.}
\label{fig4}
\end{figure}

\begin{figure}
\epsscale{1.2}
\plotone{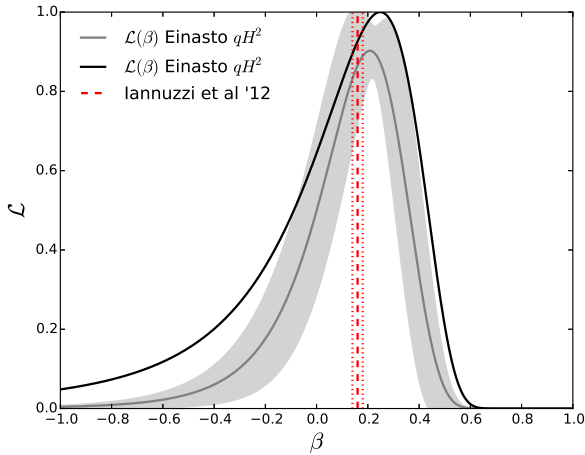}
\caption{The likelihood of the average $\beta$ as inferred from the set of 20 archival clusters (black line). We also re-plot the likelihood band (gray) from the ten sets of 20 synthetic clusters as well as the results from numerical simulations from \cite{iannuzzi} (as in Fig. \ref{fig3}). With just 20 clusters we can recover the velocity anisotropy only if the escape velocity profile is described by an Einasto density field with an additional cosmological term (as in equation \ref{eq:vesc}).}
\label{fig5}
\end{figure}

The ratio for our sample of 20 clusters, similar to Fig. \ref{fig2}, is shown in Fig. \ref{fig4}. As with the synthetic data, we calculate the $\chi^2$ and after assuming a Gaussian we can infer the likelihood for the average anisotropy parameter for our set of 20 clusters. The result is shown in Fig. \ref{fig5} where we compare the likelihood of $\beta$ inferred for the archival set of 20 clusters (black line) to the distribution of $\mathcal{\beta}$ inferred from the synthetic set of clusters (gray band, as in Fig. \ref{fig3}). 

We find that for our sample of just 20 clusters we recover the peak of the likelihood as inferred from simulations to high accuracy. In particular we find that the value of the most likely anisotropy parameter for our archival data of 20 galaxy clusters is  $ \beta = 0.248_{-0.360}^{+0.164} $ at the 68\% confidence level. Note that we calculate this interval by assuming that the distribution to the right of the peak is Gaussian and then find where $Q > -2 \ln [\mathcal{L}(\beta)/\mathcal{L}_{max}]$ (where for our single parameter model $Q= 1$). After this error is calculated then we find the error to the left of the peak by integrating the likelihood from the rightmost error up to the value of $\beta$ which yields 0.68 times the total area of the likelihood. 

In Fig. \ref{fig5} the data likelihood (black line) is larger than the simulation likelihood for $\beta$ (gray band) and with a longer tail to negative $\beta$. This is because in the simulations we applied a representative error on the cluster masses fixed at 25\%, whereas the data have errors which vary from 15\% to 44\%. The larger errors on the weak lensing masses allow for more negative anisotropy. In fact, as we discuss below, the weak lensing mass errors are the dominant component of our error budget.

\section{systematics}
The agreement between the anisotropy parameter inferred with the archival data of 20 clusters and the $\Lambda$CDM simulation results (assuming the Einasto $qH^2$ theory) is clear. To understand the robustness of this result, we carefully consider possible systematics. 

As mentioned in the previous section, we carefully consider the uncertainties that make up our averaged ratio and we weight our data with those errors. In particular, we weight the ratio for a given cluster at a given radius by its error at that radius. More specifically, we propagate the error on the ratio by considering the error on the numerator (i.e. $v_{esc}^{edge}(r)$) and the error on the denominator (i.e. the theoretical escape velocity profile, $v_{esc}(r)$). We consider various uncertainties plaguing these two components of the average ratio below.

Throughout, we use a $\chi^2$ statistic and assume that the errors on the ratios are Gaussian. The fact that our likelihood is centered near the truth for the simulated data indicates that the assumption of Gaussianity is not a bad choice for our analysis.

\subsection{Escape velocity edge}
We propagate the error on the averaged ratio by assuming the error on the numerator, that is on the edge  ($v_{esc}^{edge}$), at each cluster and at each radial bin to be 5\% (as studied by \cite{miller2016}). As mentioned previously, the edge is characterized by the galaxies with projected velocities in the top 10\% of the phase-space at a given radial bin. We test the robustness of the edge-detection technique by changing this percentage by 50\% and find that the resulting variations lie within our 68\% confidence interval error. We note that these variations on the edge affect both the Einasto and Einasto $qH^2$ $\beta$ likelihoods equally. As such, it has no direct effect on our results. Furthermore, we utilize the interloper removal technique described in \cite{gifford2013} which was tested against different cosmologies in Gifford et al. (2016) to infer the edge to high precision using $\sim 50$ galaxies per phase-space. Note that our clusters have much higher phase-space sampling and thus better contrasted edges. We also compared our edges to those measured with completely independent techniques (e.g., \cite{geller2013}) and find no significant differences within the scatter and limited sample size used. Given the expectations from theory, the tests in simulated projected data, and the comparison to other observed measurements, we do not consider this component of the average ratio calculation to be a significant source of systematic uncertainty. 

\subsection{Weak lensing masses}
\subsubsection{Scatter}
In the denominator of the ratio, we calculate the error on the theory ($v_{esc}$) by folding through the error in the weak lensing mass estimates. The mean ratio is weighted according to its total error, and so our mean ratios can vary slightly if the weak lensing mass errors change. More importantly, the likelihood will shrink (grow) as the weak lensing mass errors shrink (grow).  Nonetheless the dominant error on the ratio comes from the weak lensing mass estimates.

Lastly, note that in our calculation of the error we assume that the error on the $v_{esc}^{edge}(r)$ and the error on $v_{esc}(r)$ from the mass scatter have no covariance. This could either raise or decrease our overall error.

\subsubsection{Bias}
Another possible systematic that could affect our theoretical expectation is biased weak lensing masses. In particular, biasing all of the weak lensing masses in our archival data sample by 5\% (i.e. $M_{200} \rightarrow 0.95 \times M_{200}$), as is perhaps expected by \cite{beckerkravtsov}, would change the most likely velocity anisotropy parameter to $\beta = 0.138$. This value is still within a reasonable range of our likelihood expectation from the synthetic cluster sample. However, we note that given that we recover the results in simulations by using the same technique we can be confident that we are not utilizing biased masses in our archival data sample.

\subsection{Mass-concentration relation}
 Another component that is implicit in our calculation of $v_{esc}(r)$ is the utilization of a mass-concentration relation to attain the NFW density profiles. Recalling from previous sections, we utilize this NFW density to infer the Einasto parameters as described above. In particular, we utilize the \cite{duffy} mass-concentration relation for both the synthetic and archival data samples. Most importantly, this is the relation also used in the metacatalog  we utilize \citep{sereno}. The relation is given by,
 \begin{equation}
 c_{200}(M_{200}, z) = A_{200} \Bigg(\frac{M_{200}}{M_{piv}} \Bigg)^{B_{200}} (1+z)^{C_{200}}.
 \end{equation}
  Where $A_{200}= 5.71$, $B_{200} = -0.084$,  $C_{200} = -0.47$ and $M_{piv} = 2\times 10^{12} h^{-1} M_{\odot}$. We re-calculated our inferred $\beta$ from our samples by employing the $1\sigma$ error variations on the relation's parameters ($A_{200},B_{200}, C_{200}$) and found that the inferred $\beta$ varied only by $\sim 1\%$. Perhaps this is expected given the relative flatness of the mass-concentration relation in the high mass end of the spectrum which is where most of our clusters lie. 

\subsection{Cosmological parameters}

As expressed in the preceding sections, our theoretical expectation for the projected escape velocity profile involves assuming a cosmology. Therefore, we expect that the uncertainty in these cosmological parameters will also affect our theoretical escape velocity profile ($v_{esc}(r)$), and consequently, $\beta$. We note that the variations for these cosmological parameters are significant. This much is already implied by Fig. \ref{fig1}.

In particular we note that decreasing $\Omega_M$ to 0.01 changes the peak of the anisotropy likelihood to $\beta = -0.08$. This significant difference in $\beta$ is due to the fact that the escape velocity profiles would be underestimated in relation to the inferred escape velocity edges. Similarly, increasing $\Omega_M$ increases the escape velocity profiles and raises $\beta$, thereby shifting the peak to the right.  Picking a more realistic uncertainty, we find that a 10\% variation in $\Omega_M$, yields a 19\% variation in the peak of $\beta$.

Increasing $H_0$ for a constant $\Omega_M$ also increases $\beta$. The variations are even more accentuated. For example, a 2\% variation in $H_0$ yields a 42\% variation in the peak of $\beta$. These variations are still within the 95\% confidence region as shown in Fig. \ref{fig5}.

We emphasize that despite these significant variations on the inferred value of $\beta$ our goal in this paper is to test whether we can accurately and precisely reproduce projected escape velocity edges given a cosmology-dependent model.

\section{discussion}

By utilizing archival data of just 20 galaxy clusters and by picking a fiducial cosmology within the range of what is expected from cosmological probes, we are able to recover the average anisotropy parameter in agreement with $\Lambda$CDM simulations. In this sense, then, we are already implicitly constraining cosmology by picking a reasonable choice of values for $h$, $\Omega_m$ and $w$. What remains to be seen, however, and this much we defer to future work, is how precisely we can constrain cosmology with Eq. \ref{eq:vesc} once $\beta$ is independently inferred for each cluster given the scatter on weak lensing masses (our dominant systematic). If $\beta$ can be inferred for each cluster through an independent technique (e.g. via the Jean's equation), we can leverage this to constrain cosmological parameters in the near future. 

We note that our resulting average velocity anisotropies are well in agreement with other analyses \citep{macs1206,lemze2009,lokasabell,wojtak,benatov}. For example, with a sample of only 6 nearby relaxed Abell clusters, \cite{lokasabell} constrains the anisotropy parameter to $-1.1 < \beta < 0.5$ at the 95\% confidence level. Our results with 20 clusters basically reproduce this constraint on $\beta.$ 

Furthermore, we note that our treatment would not affect observables that are either first or second derivatives of the potential, such as Jean's mass analyses or inferences of X-ray masses. The reason for this is that the dominant term in equation \ref{eq:vesc} is the normalization constant $\Psi(r_{eq})$. The techniques that are affected by our theoretical expectations are those that directly deal with the escape velocity profile as such (or in other words, with the cluster's potential as inferred from dynamics). Therefore our analyses matters when the normalization of the potential matters. This is not the case for the two aforementioned methods nor for weak lensing masses. In short, cosmology matters for escape-velocity inferred masses, as shown in 3-dimensional $\Lambda$CDM simulation (\cite{miller2016}),  2-dimensional simulations (Fig. \ref{fig3}), and real data (Fig. \ref{fig5}, black line).

One technique that directly deals with the escape velocity profile as such is the caustic technique. \cite{hoekstra2015} has recently demonstrated that mass estimates by the caustic technique (see \cite{hecs}) are on average underestimated when compared to the weak lensing masses by $\sim 22 \%.$ If we drop the cosmological terms in our theoretical profile, the overall escape velocity profile would increase (as in Fig. \ref{fig1}). As such, in order to match this Einasto profile without the cosmological term, the Einasto $qH^2$ would always infer a higher mass. Interestingly, we find that $M_{200}$(Einasto $ qH^2) /M_{200}$(Einasto) = 1.22, exactly reproducing \cite{hoekstra2015}'s result. We note this as a possible explanation for this discrepancy and as a call to return to reflect upon the cosmological dependence of the escape velocity edges as already argued long ago by \cite{kaiser1986} and \cite{regos}.

\section{Conclusions}
With archival data of just 20 galaxy clusters with extensive redshift information and weak lensing mass profile estimates we demonstrate the need to include a cosmological dependent term in the analytic model of the escape velocity profile of galaxy clusters. We conduct our analysis also utilizing ten sets of 20 synthetic galaxy clusters to study underlying systematics and projection effects. We find that our analytic formulation provides remarkable agreement with both sets of data. 

More specifically, we leverage the complications involved in projecting the line-of-sight velocities related to to the anisotropy parameter ($\beta$) and utilize this information to quantify the necessity of including a cosmological term in our analytic theory. 

We find that if we do not include a cosmological term in our analytic theory we infer velocity anisotropies that are inconsistent with both numerical results and observational data. 

Throughout our analysis we picked a fiducial cosmology to probe our theoretical expectations and showed that there is a degeneracy between the velocity anisotropy parameter and cosmology. However, by independently inferring the anisotropy parameter, and combining this with the cosmology dependent Einasto $qH^2$ theoretical profiles, one can in principle constrain cosmology.

As such, we have briefly motivated the capacity for the escape velocity profile of galaxy clusters to become a novel  probe of cosmology in the near future given its sensitivity to the physics of cosmic acceleration. What is required to realize this is the next generation of weak-lensing data for galaxy clusters as well as deep spectroscopic follow-up (e.g., the Dark Energy Survey \cite{Diehl} and the Dark Energy Spectroscopic Instrument--DESI). We defer a systematic of study of this kind to future work.

\section{acknowledgments}

 This material is based upon work supported by the National Science Foundation under Grant No. 1311820 and 1256260. The Millennium Simulation databases used in this paper and the web application providing online access to them were constructed as part of the activities of the German Astrophysical Virtual Observatory (GAVO). This research has made use of the VizieR catalogue access tool, CDS, Strasbourg, France. Lastly, we would like to thank the anonymous referee whose comments have improved both the quality of presentation and level of rigor of our analysis.

\bibliographystyle{apj}
\bibliography{stark}
\newpage

\end{document}